\documentclass{emulateapj}

\def\ptf{PTF~10qpf}
\def\lkh{LkH$\alpha$ 188-G4}
\def\kms{km~s$^{-1}$}

\def\arcmin{\hbox{$^\prime$}}
\def\arcsec{\hbox{$^{\prime\prime}$}}

\def\lsun{L$_\odot$}

\def\jhk{{\em J}, {\em H}, and {\em K$_s$}}

\def\ebv{$E(B$--$V)$}

\usepackage{natbib,epsfig,graphicx,rotating,lscape,rotating}
\slugcomment{DRAFT \today}
\shorttitle{A New FU Orionis Outburst From a CTTS}
\shortauthors{Miller et.\ al.}

\begin{document}

\title{Evidence for an FU Orionis-like Outburst from a Classical T Tauri Star}

\author{Adam~A.~Miller\altaffilmark{1}, 
Lynne~A.~Hillenbrand\altaffilmark{2}, 
Kevin~R.~Covey\altaffilmark{3,4,5}, 
Dovi Poznanski\altaffilmark{1,6,7},
Jeffrey M. Silverman\altaffilmark{1},
Io K. W. Kleiser\altaffilmark{1},
B\'{a}rbara Rojas-Ayala\altaffilmark{3},
Philip S. Muirhead\altaffilmark{3},
S. Bradley Cenko\altaffilmark{1},
Joshua S. Bloom\altaffilmark{1},
Mansi M. Kasliwal\altaffilmark{2},
Alexei V. Filippenko\altaffilmark{1},
Nicholas M. Law\altaffilmark{8},
Eran O. Ofek\altaffilmark{2,7},
Richard G. Dekany\altaffilmark{9}, 
Gustavo Rahmer\altaffilmark{9}, 
David Hale\altaffilmark{9}, 
Roger Smith\altaffilmark{9}, 
Robert M. Quimby\altaffilmark{2}, 
Peter Nugent\altaffilmark{6}, 
Janet Jacobsen\altaffilmark{6}, 
Jeff Zolkower\altaffilmark{9}, 
Viswa Velur\altaffilmark{9}, 
Richard Walters\altaffilmark{9}, 
John Henning\altaffilmark{9}, 
Khanh Bui\altaffilmark{9}, 
Dan McKenna\altaffilmark{9}, 
Shrinivas R. Kulkarni\altaffilmark{2}, 
and Christopher R. Klein\altaffilmark{1}
}

\altaffiltext{1}{Department of Astronomy, University of California, Berkeley, CA 94720-3411, USA.}
\altaffiltext{2}{Astrophysics Department, California Institute of Technology, Pasadena, CA 91125, USA.}
\altaffiltext{3}{Department of Astronomy, Cornell University, Ithaca, NY 14853, USA.} 
\altaffiltext{4}{Hubble Fellow.} 
\altaffiltext{5}{Visiting Researcher, Department of Astronomy, Boston University, 725 Commonwealth Ave, Boston, MA 02215, USA.} 
\altaffiltext{6}{Computational Cosmology Center, Lawrence Berkeley National Laboratory, 1 Cyclotron Road, Berkeley, CA 94720, USA.} 
\altaffiltext{7}{Einstein Fellow.} 
\altaffiltext{8}{Dunlap Institute for Astronomy and Astrophysics, University of Toronto, 50 St. George Street, Toronto M5S 3H4, Ontario, Canada.} 
\altaffiltext{9}{Caltech Optical Observatories, California Institute of Technology, Pasadena, CA 91125, USA.}

\begin{abstract}
We present pre- and post-outburst observations of the new FU
Orionis-like young stellar object \ptf\ (also known as \lkh\ and
HBC 722). Prior to this outburst, \lkh\ was classified as a
classical T Tauri star on the basis of its optical emission-line
spectrum superposed on a K8-type photosphere, and its photometric
variability. The mid-infrared spectral index of \lkh\ indicates a
Class II-type object. \lkh\ exhibited a steady rise by $\sim$1 mag
over $\sim$11 months starting in Aug.\ 2009, before a subsequent more
abrupt rise of $>$ 3 mag on a time scale of $\sim$2 months.
Observations taken during the eruption exhibit the defining
characteristics of FU Orionis variables: (i) an increase in
brightness by $\ga$4 mag, (ii) a bright optical/near-infrared
reflection nebula appeared, (iii) optical spectra are consistent with
a G supergiant and dominated by absorption lines, the only exception
being H$\alpha$ which is characterized by a P Cygni profile, (iv)
near-infrared spectra resemble those of late K--M giants/supergiants
with enhanced absorption seen in the molecular bands of CO and H$_2$O,
and (v) outflow signatures in H and He 
are seen in the form of blueshifted absorption profiles.  \lkh\ is
the first member of the FU Orionis-like class with a well-sampled optical
to mid-infrared spectral energy distribution in the pre-outburst
phase.  The association of the \ptf\ outburst with the previously
identified classical T Tauri star \lkh\ (HBC 722) provides strong
evidence that FU Orionis-like eruptions represent periods of enhanced disk
accretion and outflow, likely triggered by instabilities in the disk. 
The early identification of \ptf\ as an FU Orionis-like variable will 
enable detailed photometric and spectroscopic observations during its 
post-outburst evolution for comparison with other known outbursting objects.
\end{abstract}


\section{Introduction}

FU Orionis variables are young stellar objects (YSOs) that exhibit
large-amplitude optical outbursts ($\Delta m_{\rm visual} \ga 4$ mag)
and remain bright for several decades \citep{herbig77}. Models in
which the radiation is dominated by a rapid increase in the disk
accretion rate have successfully reproduced many of the observational
properties of FU Orionis outbursts; see \citet{hartmann96} for a
review. An alternative interpretation involving a rapidly rotating
low-gravity star is presented by \citet{herbig89} and
\citet{herbig03}. In the first scenario, the cause of the instability
remains uncertain; possible mechanisms include thermal instabilities 
\citep{hartmann96} and interactions with companion stars on eccentric 
orbits \citep{bonnell92}.
Accretion instabilities are also believed to give rise to less
dramatic YSO flares.\footnote{In the literature, these events are 
frequently referred to as ``EXors'', named for the outbursting star EX 
Lupi. This nomenclature is confusing, however, as EX Lupi is not part  
of the Orion constellation and many of the so called EXors 
seem to have distinct properties (see e.g., \citealt{fedele07},  
\citealt{herbig07}, \citealt{covey11}).} These events (e.g., EX Lupi, 
V1647 Ori, V1118 Ori, Z CMa; see
\citealt{herbig77}, \citealt{herbig07}, \citealt{aspin10},
\citealt{aspin09}, \citealt{fedele07}, \citealt{rettig05},
\citealt{lorenzetti07}, \citealt{audard10}, \citealt{szeifert10}, and
references therein) are heterogeneous, and it remains to be determined
if they should be referred to as a single class
\citep{herbig07}. These YSO outbursts are nevertheless
distinguished from FU Orionis-like outbursts in that their optical
spectra are emission-line dominated (whereas FU Ori stars typically
show only absorption lines during outburst) and the outbursts 
have shorter lifetimes, $\tau_{\rm outburst} \la$ 2 yr, 
than FU Ori stars which remain in ourburst for decades 
\citep{herbig77, fedele07}.

YSO outbursts are of great interest given their potential importance
to the broader star and planet formation process.  Low-mass stars
could accrete as much as half their final mass during FU Orionis
outbursts \citep{hartmann96}, and the strong winds and outflows they
are thought to drive likely have a significant impact on the
surrounding interstellar medium \citep{croswell87}.

One of the major challenges in studying FU Orionis stars is the
relatively small sample of known examples. Although $\sim$20 FU
Orionis candidates have been identified by \citet{reipurth10}
based on a mix of spectroscopic evidence, only a handful ($\sim$6--8)
of those stars have been observed to rise from their pre-outburst
state to their eruptive state. Only one example, V1057 Cyg, has had a
pre-outburst spectrum taken; it showed Balmer, Ca~II, Fe~I, and Fe~II
lines in emission \citep{herbig77}. Herbig noted that these features
resembled those of a T Tauri star, but a lack of detected absorption
lines precluded a spectral-type classification beyond a late-type star
based on the red continuum.

Recently, a new FU Orionis-like variable, associated with \lkh, was
discovered at $\alpha_{\rm J2000} = $ 20$^{\rm h}$58$^{\rm
  m}$17$\fs$00, $\delta_{\rm J2000} = $ +43$^{\rm
  o}$53\arcmin42\farcs9 in the direction of NGC 7000/IC 5070, also
known as the North America/Pelican Nebula. The source was announced as
a likely FU Ori-like outburst on 2010 Aug.\ 17 UT\footnote{UT 
dates are used throughout unless otherwise noted.} by \citet{semkov10},
and was later spectroscopically confirmed as an FU Ori candidate by
\citet{munari10}. Further photometric observations showed that the
source became bluer over the course of the outburst and that a
reflection nebula was emerging around \lkh\ \citep{semkov10a}.
Near-infrared (NIR) photometric observations revealed a $\sim$3 mag
increase in the \jhk\ bands \citep{leoni10}. The variable was
independently discovered by our collaboration during the course of
regular monitoring of the North America Nebula with the Palomar 48-in
telescope. These images were automatically reduced, and new variable
and transient sources were found using the software developed for the
Palomar Transient Factory (PTF; \citealt{law09, rau09}; Bloom et
al.\ 2011, in prep.). On 2010 Aug.\ 5 the outburst from \lkh\ was
identified and given the name \ptf.

Here, we present pre- and post-eruption observations of \lkh, an 
FU Orionis-like star, which we argue was a classical 
T Tauri star prior to eruption. Shortly before the submission of this
article, an independent analysis of \lkh\ was posted on the arXiv by
\citet{semkov10c}, in which they also argue that the 2010 eruption
from \lkh\ is an FU Orionis outburst. In \S~\ref{sec:pre-obs} we
discuss archival observations of \lkh\ prior to the 2010 eruption.
Our new 2009 and 2010 observations are presented in \S~\ref{sec:obs},
and \S~\ref{sec:analysis} contains our analysis of those
observations. We discuss our conclusions in
\S~\ref{sec:conclusions}. Throughout this paper we designate the 2010
outburst as \ptf, while we refer to the pre-outburst star as \lkh. 

\section{Pre-Outburst Observations of \lkh}\label{sec:pre-obs}

\ptf\ is associated with the previously known optical source
\lkh\ \citep{cohen79}, which was later cataloged by \citet{herbig88}
as HBC 722.  \citet{cohen79} identified the star as an $m_V=18.9$ mag
emission-line object with line equivalent widths of EW$_{\rm
  {H}\alpha} = -100.5$~\AA\ and EW$_{{\rm H}\beta} = -25.8$~\AA; [O~I]
$\lambda$6300 was also present, and NIR photometry was reported.  We
are unable to identify the strong emission feature near 4640~\AA\ in
the \citet{cohen79} spectrum, and the authors themselves do not
measure or comment on the feature.  It could be related to multiply
ionized C or N, or He~II (see, e.g., \citealt{sargent91}), though the
absence of a known exact wavelength for the feature precludes a secure
identification. The original spectrum is no longer available
(M.\ Cohen, 2010, private communication) and we caution that the
feature may be an error in the original data.  \citet{cohen79}
determined a spectral type of K7--M0; they also derived an extinction
estimate of $A_V=3.4\pm 1.2$ mag and a bolometric luminosity of
log$(L/{\rm L}_\odot) = 0.43$.

The source is part of a linear chain of young stellar objects that was
first designated by \citet{cohen79} as the NGC 7000 / IC 5070
LkH$\alpha$ 188 group. {\it Spitzer} IRAC and MIPS observations
presented by \citet{guieu09} and \citet{rebull11} later revealed that
the optically visible stars are only the surface population of a much
richer embedded cluster, known as the ``Gulf of Mexico'' cluster due
to its location relative to the North America Nebula. The Two Micron
All Sky Survey (2MASS; \citealt{skrutskie-2mass}) observed the field
in the \jhk\ bands on 2000 June\ 10. $BVI$ optical photometry obtained
with the Kitt Peak National Observatory 0.9-m telescope was presented
by \citet{guieu09}, while broadband $ri$ and narrowband H$\alpha$
imaging of \lkh, obtained between 2003 and 2005, is presented in the
catalog of the INT/WFC Photometric H$\alpha$ Survey of the Northern
Galactic Plane (IPHAS; \citealt{gonzalez-solares08}). We summarize the
IPHAS measurements in Table~\ref{tab-iphas}.

Owing to crowding, there are no {\it IRAS} measurements of
\lkh. Additional IR sources are cataloged by the Midcourse Space
Experiment ({\it MSX}; \citealt{mill94}) and {\it AKARI}
\citep{murakami07}. The {\it MSX} source has a position ($\alpha_{\rm
  J2000} = $ 20$^{\rm h}$58$^{\rm m}$17$\fs$59, $\delta_{\rm J2000} =
$ 43$^{\rm o}$53\arcmin36\farcs24) which is $\sim$9.2\arcsec\ from the
optically derived position of \lkh. This position has no clear
counterpart, to within $\sim$3\arcsec, in the 2MASS catalog, and thus
the {\it MSX} source may represent a blend of emission from several
stars in the LkH$\alpha$ 188 group. There are two sources in the 
{\it AKARI} MIR catalog \citep{ishihara10} within 10\arcsec\ of the
optical position of \lkh. Both of these sources are
$>$4.5\arcsec\ from the optical position of \lkh, however, so we do
not include them here in our analysis of pre-outburst emission from
\lkh; the photometry of these sources likely does not correspond to
\lkh\ alone. {\it AKARI} also catalogs a source in the far-infrared
(FIR) at $\alpha_{\rm J2000} = $ 20$^{\rm h}$58$^{\rm m}$16$\fs$809,
$\delta_{\rm J2000} = $ +43$^{\rm o}$53\arcmin41\farcs65, which is
$\sim$2.4\arcsec\ from the optical position of \lkh.
The closest 2MASS counterpart to the {\it AKARI} FIR source is \lkh,
though we cannot rule out the possibility that the {\it AKARI} flux
measurements include a contribution from other members of the
LkH$\alpha$ 188 group.

\begin{figure}
\begin{center}
%\epsscale{0.9}
\includegraphics[width=85mm]{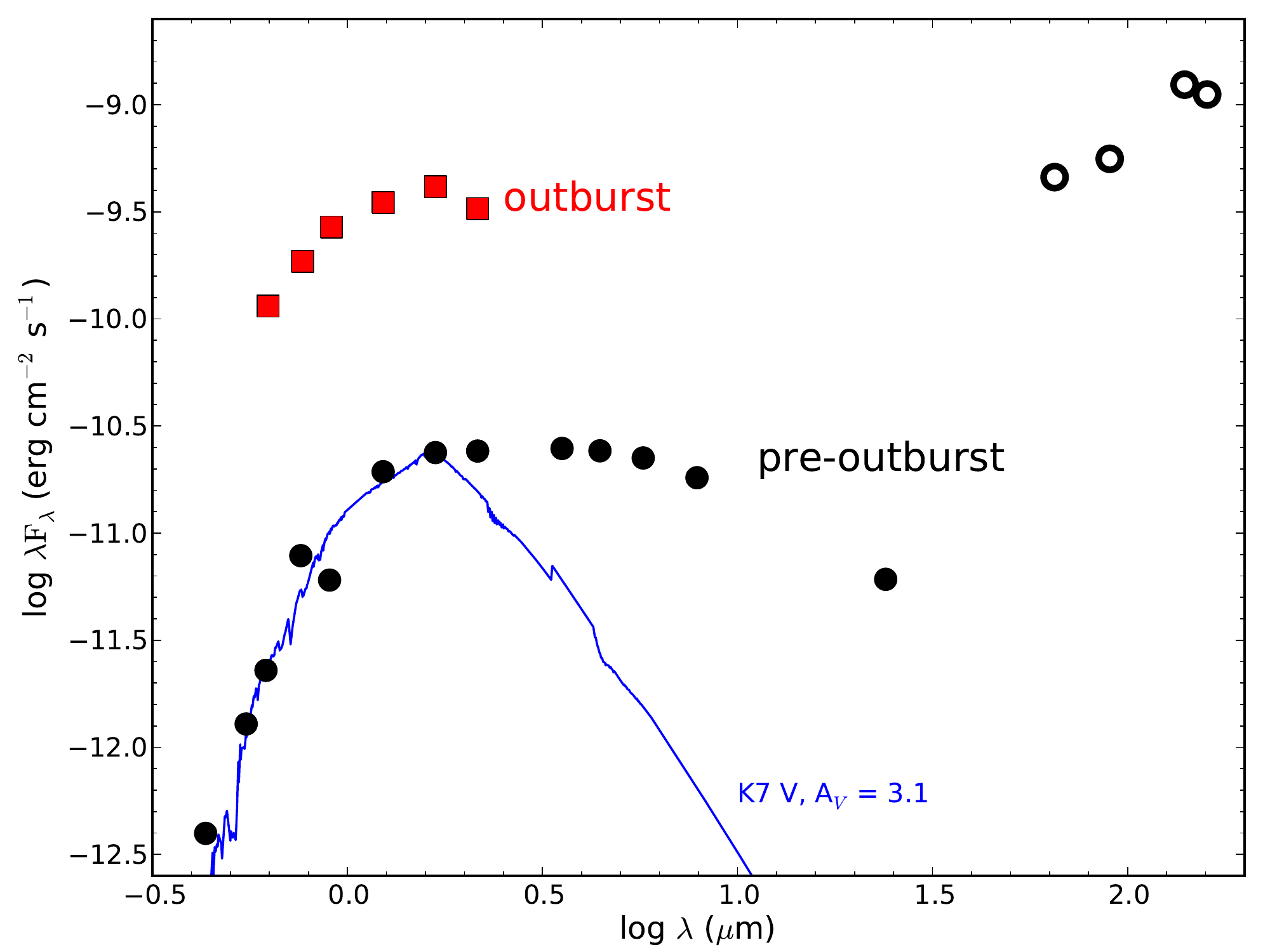}
\caption{Spectral energy distribution of \lkh\ both before (black
  circles) and during (red squares) the 2010
  outburst. Pre-outburst NIR photometry is from 2MASS, while non-simultaneous 	
  optical and MIR photometry is from \citet{guieu09},
  \citet{rebull11}, and \citet{gonzalez-solares08}. FIR photometry from
        {\it AKARI} is shown as open circles because these detections
        may include light from other members of the LkH$\alpha$ 188 group
        (see text and figure~\ref{reflect-nebula}). The solid line shows 
	the model spectrum, from
        \citet{kurucz93}, of a K7~V star reddened with $A_V$ = 3.1
        mag, and demonstrates the infrared excess of \lkh. The
        spectral slope in the MIR is consistent with the source being
        a Class II YSO prior to outburst.  NIR outburst photometry
        was obtained on 2010 Oct.\ 10 with PAIRITEL, while $riz$
        photometry was taken on 2010 Oct.\ 14 with P60.  }
\label{progSED}
\end{center}
\end{figure}

The pre-outburst spectral energy distribution (SED) of \lkh\ is shown
in Figure~\ref{progSED}. The pre-outburst SED of
\lkh\ exhibits a smooth trend with the exception of the two points at
$\sim$0.75 and 0.9 $\mu$m. These points, corresponding to the IPHAS
$i$ band and KPNO $I$ band, respectively, are not coeval and differ
by a factor of $\sim$30\% in flux, which is similar to the $\sim$0.3
mag variability seen in the multi-epoch IPHAS data.

Direct integration of the pre-outburst SED, excluding the {\it AKARI}
data, results in a total luminosity of $\sim$0.7 \lsun, assuming a
distance of 600 pc to NGC 7000.  Adopting the classification scheme
from \citet{lada84}, we fit for the MIR spectral index, defined as
\begin{equation}
	\alpha = \frac{d \log \lambda F_\lambda}{d \log \lambda},
\end{equation}
% %
where $\alpha$ is the spectral index and $F_\lambda$ is the flux
density at wavelength $\lambda$. Following a least-squares fit to the
{\it Spitzer}/IRAC data from 3.6 to 8 $\mu$m, we measure $\alpha =
-0.40$, indicating that \lkh\ was a Class II YSO {\it prior} to its
2010 outburst. Thus, there was no evidence of either a dense core of
infalling material or significant envelope material in the vicinity of
the central star. Including the {\it Spitzer}/MIPS 24 $\mu$m detection
results in $\alpha = -0.77$, securely establishing this source as a
Class II YSO. To demonstrate the MIR excess relative to a stellar
photosphere, we also show in Figure~\ref{progSED} the model spectrum of
a solar metallicity K7~V star, reddened with $A_V = 3.1$ mag, which is
consistent with the inferences from \citet{cohen79}.  The model
spectrum, taken from the \citet{kurucz93} update to the
\citet{kurucz79} models, has been normalized to match the flux of
\lkh\ in the $H$ band.

Class II YSOs are most commonly associated with classical T Tauri
stars (CTTSs; \citealt{lada87}). In addition to being a Class II YSO,
we also know from the multi-epoch IPHAS data that \lkh\ was optically
variable between 2003 and 2005, years before the observed PTF
outburst, with an amplitude of $\sim$0.2--0.3 mag. While some CTTSs
show variable amplitudes as large as $\sim$2 mag, the majority exhibit
low-amplitude variability consistent with what we observe in
\lkh\ prior to eruption \citep{grankin07}. The emission-line spectrum,
Class II MIR spectral index, and optical variability all point to
\lkh\ being a fairly undistinguished CTTS prior to its 2010 eruption.

\section{2009 and 2010 Observations}\label{sec:obs}

\subsection{Optical Photometry and an Independent Discovery}

\begin{figure}
\begin{center}
%\epsscale{0.9}
\includegraphics[width=85mm]{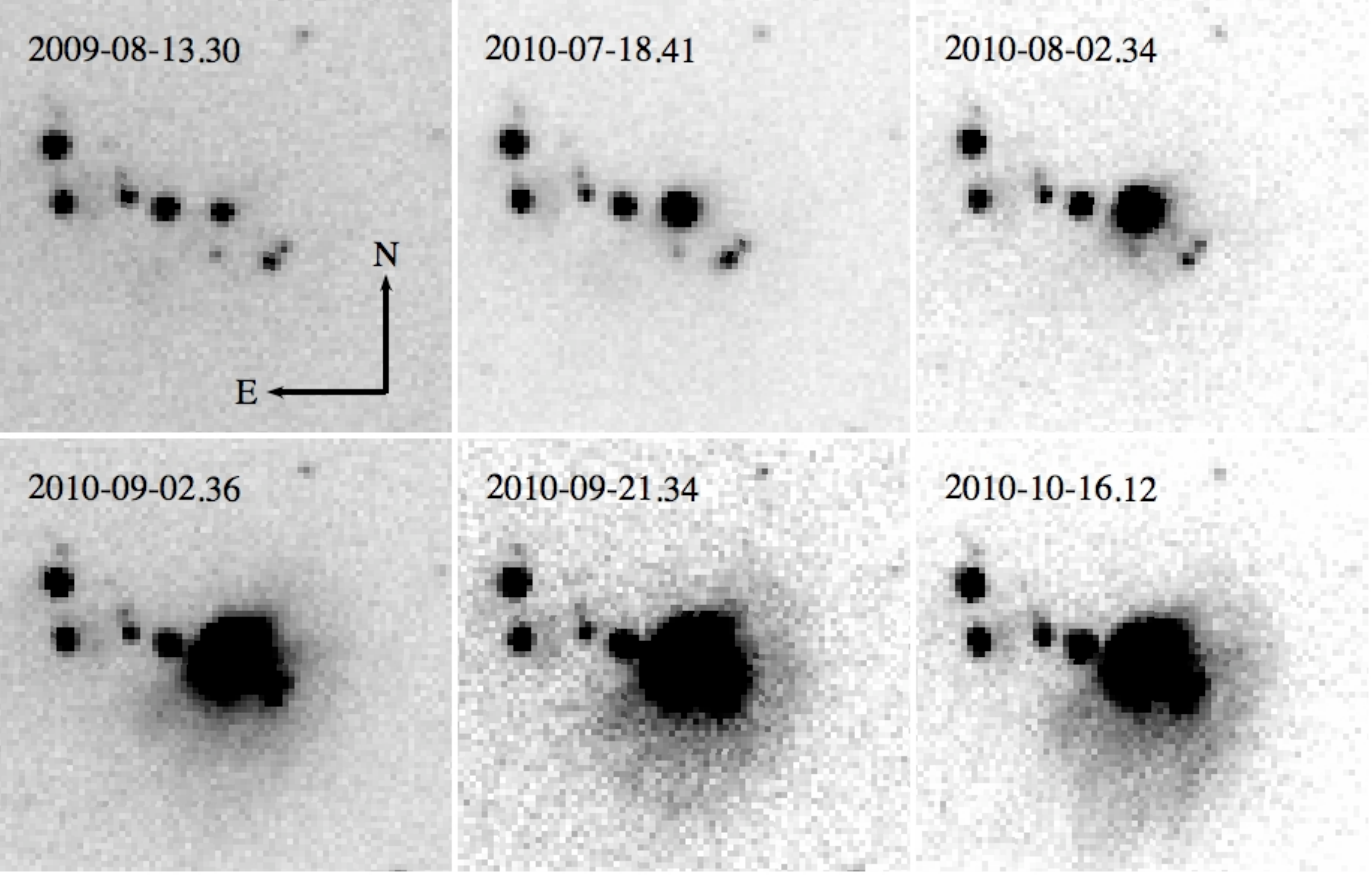}
\caption{P48 $R$-band images of \ptf\ showing the emergence of the
  optical reflection nebula. Each image is centered on \ptf\ and
  is 75\arcsec\ on a side. As with many other FU Orionis-like variables,
  the outburst of \ptf\ has created an asymmetric reflection
  nebula. Notice that \ptf\ is not located in the center of the
  reflection nebula which extends to the south west of the pre-outburst 
  point source. All images are registered to the same frame with
  north up and east to the left. All dates are UT.  }
\label{reflect-nebula}
\end{center}
\end{figure}

During the 2009 and 2010 observing seasons, PTF obtained red optical
images of the North America/Pelican Nebula star-forming region with a
typical 5-day cadence.  These observations were conducted with the
main PTF Survey Camera, the former CFHT12K mosaic camera now
extensively re-engineered and mounted on the 48-in Samuel Oschin
telescope at Palomar Observatory (hereafter P48). The camera is a
mosaic of 12 CCDs (one of which is not functional), covering a 7.8
square degree field of view with 1\arcsec\ sampling; typical
conditions at Palomar Observatory produce 2.0\arcsec\ full width at
half-maximum intensity (FWHM) images \citep{law09}.  The $R_{\rm PTF}$
filter, a Mould $R$ band, is similar to the Sloan Digital Sky Survey
(SDSS) $r$ band in shape but is shifted redward by $\sim 27$~\AA\ and
is $\sim 20$~\AA\ wider.  The typical 5$\sigma$ limiting magnitudes
are $m_{R} \approx 20.6$ (AB) in 60~s exposures.  Representative
images of \lkh/\ptf\ from the 2009 and 2010 observing seasons are
shown in Figure \ref{reflect-nebula}.

Transient sources are detected in the PTF monitoring data by means of
automated reduction pipelines, including a nearly real-time
image-subtraction pipeline hosted at Lawrence Berkeley National
Laboratory (LBNL).  Well-detected sources in the difference images are
scored (using a human-trained, machine-based classifier) for their
likelihood of being truly astrophysical in nature or of spurious
origin.  Variable sources with larger likelihoods of being nonspurious
are passed to an automatic source classifier at UC Berkeley ({\it
  Oarical}), which combines PTF measurements with all other available
information (e.g., SIMBAD identifications, 2MASS photometry, etc.) to
provide probabilistic classifications of PTF detections (Bloom et al.,
2011, in prep.).  These initial classifications are made available to
PTF collaboration members via the PTF Follow-up Marshal, which enables
visual inspection of current and reference images, precursor PTF light
curves, and any subsequent spectroscopy.

\begin{figure}
\begin{center}
%\epsscale{0.9}
\includegraphics[width=85mm]{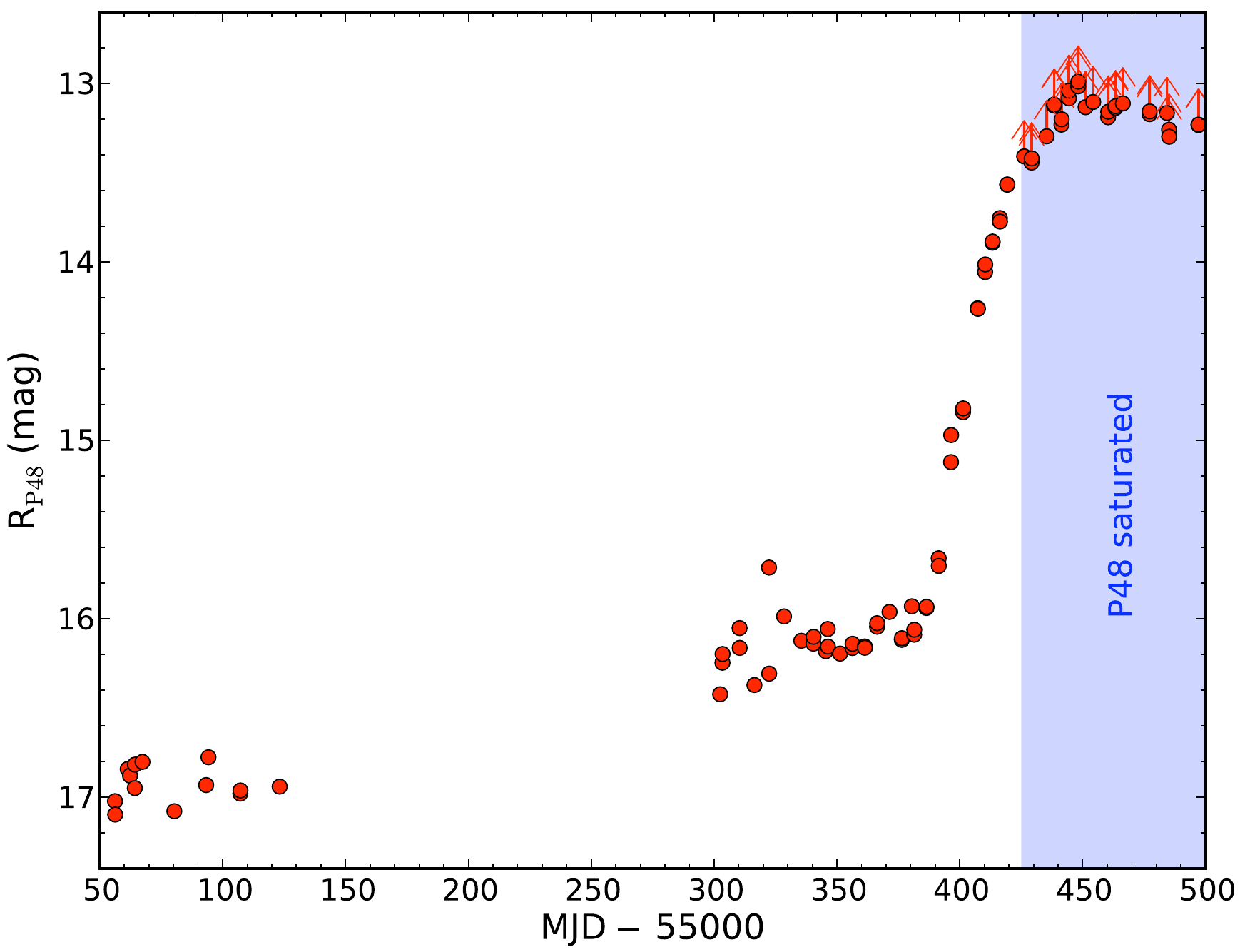}
\caption{P48 light curve showing the outburst of \lkh/\ptf. During the
  2009 season \ptf\ was variable with typical ${\Delta}m$ $\approx$
  0.2 mag. The source dramatically increased the rate at which it was
  brightening around 2010 July 13 (day $\sim$391 in the figure). P48
  saturates on sources brighter than $m_{R_{\rm P48}} \approx$ 13.5
  mag. The shaded region shows all data where \ptf\ is
  saturated; these data represent {\it lower} limits to the true
  brightness of \ptf\ during those epochs.  }
\label{P48-lc}
\end{center}
\end{figure}

The source reported here was detected by the PTF pipeline and
automatically assigned the name \ptf, following internal PTF naming
conventions. Differential photometric measurements are made via
aperture photometry, with the aperture size scaled to the seeing in
each image, using SExtractor \citep{bertin96}, and absolute
calibration is relative to the USNO-B1 catalog \citep{monet03} with an
uncertainty of $\sim 0.15$ mag. CCD 3 on the P48 camera saturates at
55,000 counts, roughly corresponding to sources with $R \la 13.5$ mag,
given the standard 60~s exposure time. Thus, in all of our
observations following those on 2010 Aug.\ 14, \lkh\ is saturated and
our measurements represent a lower limit to the true brightness of the
source. These limits are reflected in the full P48 light curve of
\lkh, shown in Figure \ref{P48-lc}, and the photometry is reported in
Table \ref{tab-P48}. We do not include the systematic $\sim 0.15$ mag
calibration uncertainty in this table because this uncertainty
represents a constant offset that would be applied to all the data in
the same way.

\begin{figure}
\begin{center}
%\epsscale{0.9}
\includegraphics[width=50mm]{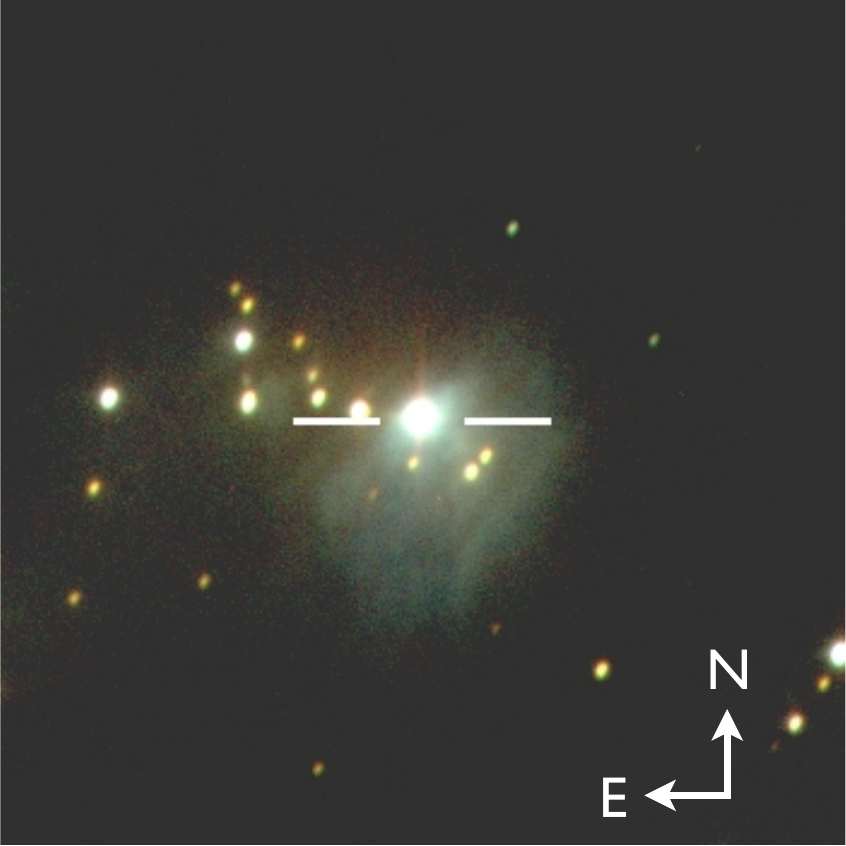}
\caption{False-color $riz$ image of \ptf\ during the 2010
  eruption. The image is 2.2\arcmin\ on a side with north up and east
  to the left. \ptf\ is highlighted with crosshairs. Notice the bright
  reflection nebula to the southwest of \ptf; this nebula was not
  present prior to the 2010 eruption.}
\label{fig:P60_frame}
\end{center}
\end{figure}

The PTF also makes use of the robotic Palomar 60-in telescope
\citep[P60;][]{cenko06} to obtain multi-filter photometry for source
verification and classification purposes.  \ptf\ was observed with the
P60 on 2010 Oct.\ 14; Figure \ref{fig:P60_frame} presents a
three-color image constructed from the P60 $riz$
frames. Table~\ref{tab-P60} contains the individual photometric
measurements. The absolute zeropoint calibration was done relative to
SDSS photometry \citep{adelman08} based on other fields observed by
the P60 during the same night with the same filter. The uncertainty in
the zeropoint is computed as the standard deviation of the zeropoints
in all other SDSS fields.

\subsection{Near-Infrared Photometry}\label{sec:NIR_phot}

Near-infrared observations of \ptf\ were conducted with the 1.3~m
Peters Automated Infrared Imaging Telescope \citep[PAIRITEL;
][]{bloom06} on Mt.~Hopkins, AZ, starting on 2010 Sep.\ 27. PAIRITEL
is a roboticized system using the former 2MASS southern hemisphere
survey camera that employs two dichroics to observe simultaneously in
the $J$, $H$, and $K_s$ bands. Observations were scheduled and
executed via a robotic system. PAIRITEL is operated in a fixed
observing mode in which 7.8 s double-correlated “images” are created
from the difference of a 7.851 s and a 51 ms integration taken in
rapid succession (see \citealt{blake08}). The standard observing
procedure involves taking three image pairs prior to dithering the
telescope.

\begin{figure}
\begin{center}
%\epsscale{0.9}
\includegraphics[width=50mm]{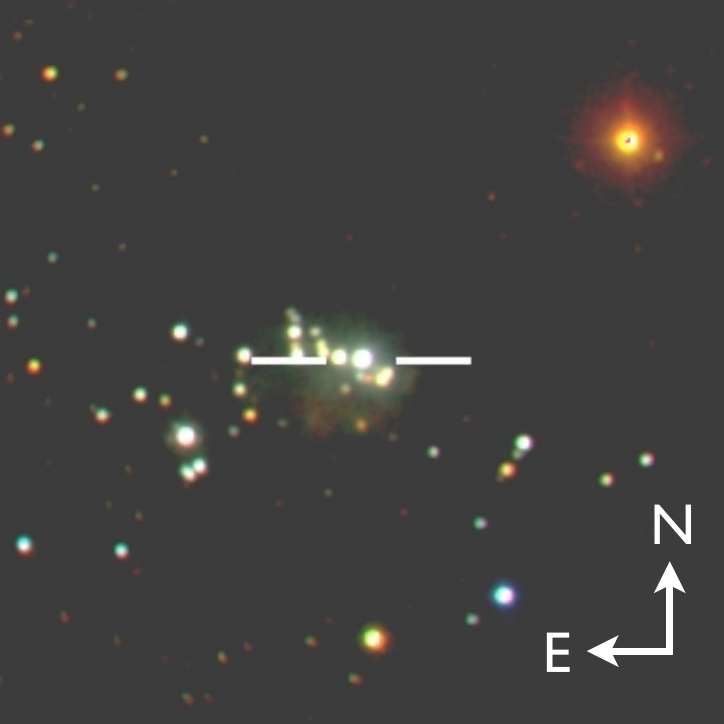}
\caption{False-color $JHK_s$ image of \ptf\ during the 2010
  eruption. The NIR reflection nebula is less prominent than the optical 
  nebula, but still visible to the southwest of \ptf. The image is 
  5\arcmin\ on a side with north up and east to
  the left. \ptf\ is highlighted with crosshairs.  }
\label{NIR-nebula}
\end{center}
\end{figure}

The raw data from these images are reduced using standard IR reduction
methods via \hbox{PAIRITEL} PIPELINE III and the flux for all sources
is measured via aperture photometry using SExtractor \citep{bertin96},
calibrated against 2MASS. In the $H$ and $K_s$ bands, \ptf\ saturates
the 7.851 s frames; however, PIPELINE III produces ``short-frame''
mosaics consisting of reduced, stacked 51 ms images (see also
\citealt{bloom080319b}). The ``short-frame'' mosaics contain $>$10
bright 2MASS stars which we use to properly calibrate photometric
measurements of \ptf\ in these images. PAIRITEL has a systematic
uncertainty of $\sim$0.02--0.03 mag in each of the \jhk\ bands (see
\citealt{blake08,perley10}), which, in the case of \lkh, is larger
than the statistical error in all three bands. Thus, we add a
systematic error of 0.03 mag in quadrature with the statistical
uncertainty to determine the total uncertainty in each band. PAIRITEL
photometry is reported in Table~\ref{tab-ptel}, and a \jhk\ band 
false-color image of \ptf\ is shown in Figure~\ref{NIR-nebula}.

\subsection{Optical Spectroscopy}

\subsubsection{Lick Spectroscopy}

\begin{figure*}
\begin{center}
%\epsscale{0.9}
\includegraphics[width=165mm]{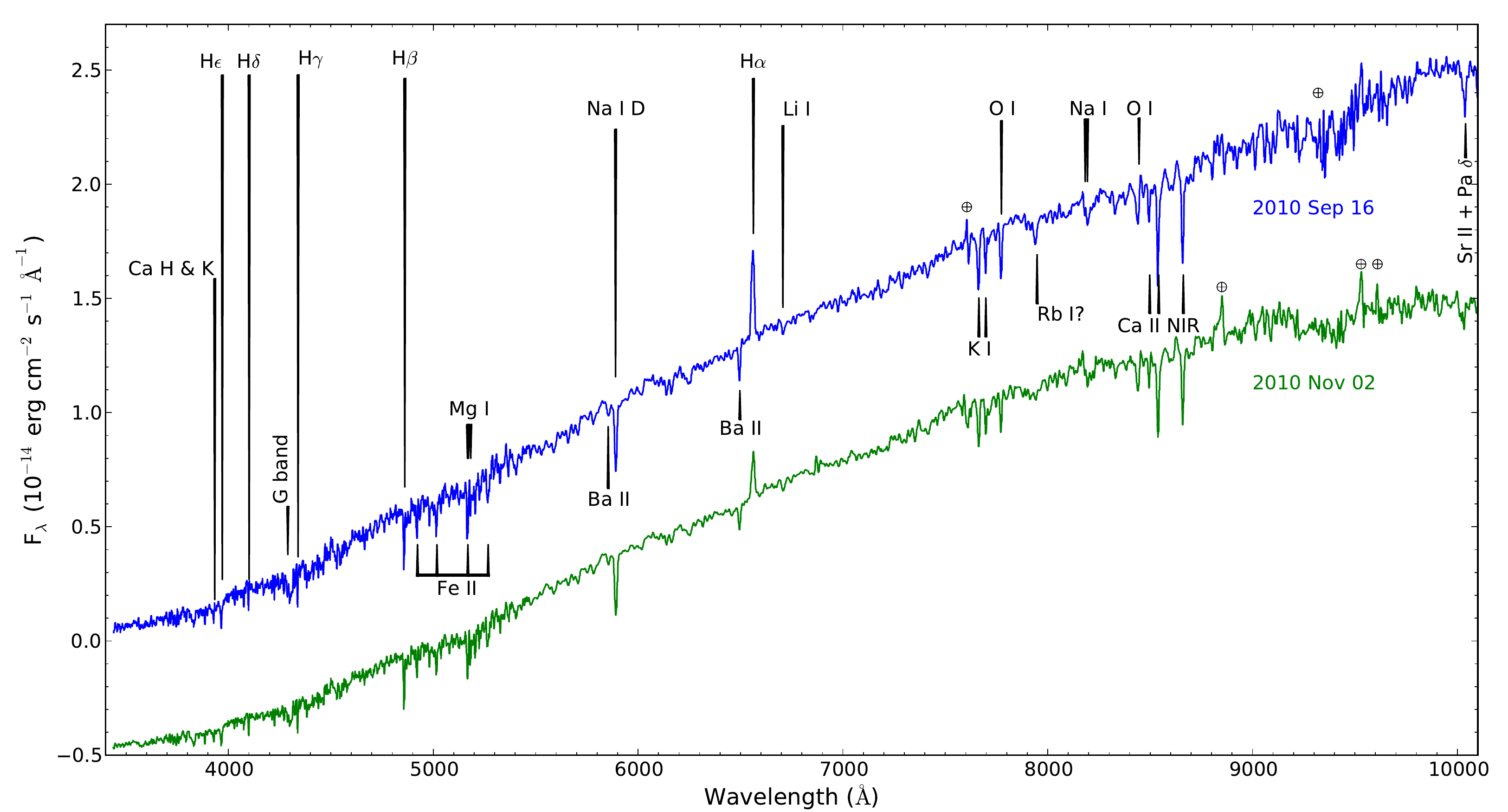}
\caption{Low-resolution spectra of \ptf. Prominent absorption lines
  are marked. H$\alpha$ is the only line seen in emission, and it
  appears to have decreased in strength between our two spectra.
  Residuals left following an imperfect removal of telluric absorption
  have been marked with a $\oplus$, while the noisy features around
  $\sim$9300~\AA\ are also likely telluric (see text). The spectrum
  from 2010 Nov.\ 2 has been shifted down by 0.5 for clarity.  }
\label{spectrum}
\end{center}
\end{figure*}

Low-resolution spectra of \ptf\ were obtained on 2010 Sep.\ 16 and
2010 Nov.\ 2 with the Kast spectrograph on the Lick 3-m Shane
telescope (Miller \& Stone 1993).  The spectra were reduced and
calibrated using standard procedures (e.g., \citealt{matheson00}). For
the Sep.\ 16 spectrum, flux calibration for the red arm of the Kast
spectrograph was performed using a standard star observed at high
airmass, whereas \ptf\ was observed at low airmass; thus, the absolute
flux calibration is somewhat uncertain in the red portion of the
optical spectrum. Furthermore, the noisy features near $\sim$9300
\AA\ are likely the result of an imperfect telluric correction and not
astrophysical. Since conditions were not photometric on the night of
Nov.\ 2, the absolute flux calibration is again somewhat
uncertain. Nevertheless, on both nights we observed \ptf\ with the
slit placed at the parallactic angle, so the relative spectral shapes
should be accurate. The full low-resolution
spectra of \ptf\ are shown in Figure~\ref{spectrum}, while all of our
spectroscopic observations are logged in Table~\ref{speclog}.

\subsubsection{High-Resolution Spectroscopy}

\ptf\ was observed on 2010 Sep.\ 25 with the Keck I telescope and
HIRES spectrometer \citep{vogt94} using the red cross-disperser and
the standard settings of the California Planet Survey \citep{howard10,
  johnson10}. The C2 decker was employed, projecting to 0\farcs86 on
the sky and providing a resolution $R = \lambda/\Delta\lambda =
55,000$ at 5500~\AA. We took advantage of the standard CPS reduction
pipeline, which includes flat-fielding, scattered-light subtraction,
order tracing, cosmic ray rejection, and spectrum extraction. After
extraction, each order is sky-subtracted and summed in the
cross-dispersion direction to form the final one-dimensional
spectrum. The exposure time of 560~s resulted in a signal-to-noise
ratio ($S/N$) of $\sim 20$ at 7400~\AA.

\subsubsection{Near-Infrared Spectroscopy}

A NIR spectrum of \ptf\ was obtained with the TripleSpec spectrograph
on the Palomar 5-m Hale telescope \citep{herter08} on 2010
Sep.\ 23. TripleSpec has no moving parts and simultaneously acquires 5
cross-dispersed orders covering 1.0--2.4 $\mu$m at $R \approx 2700$.

\begin{figure*}
\begin{center}
%\epsscale{0.9}
\includegraphics[width=140mm]{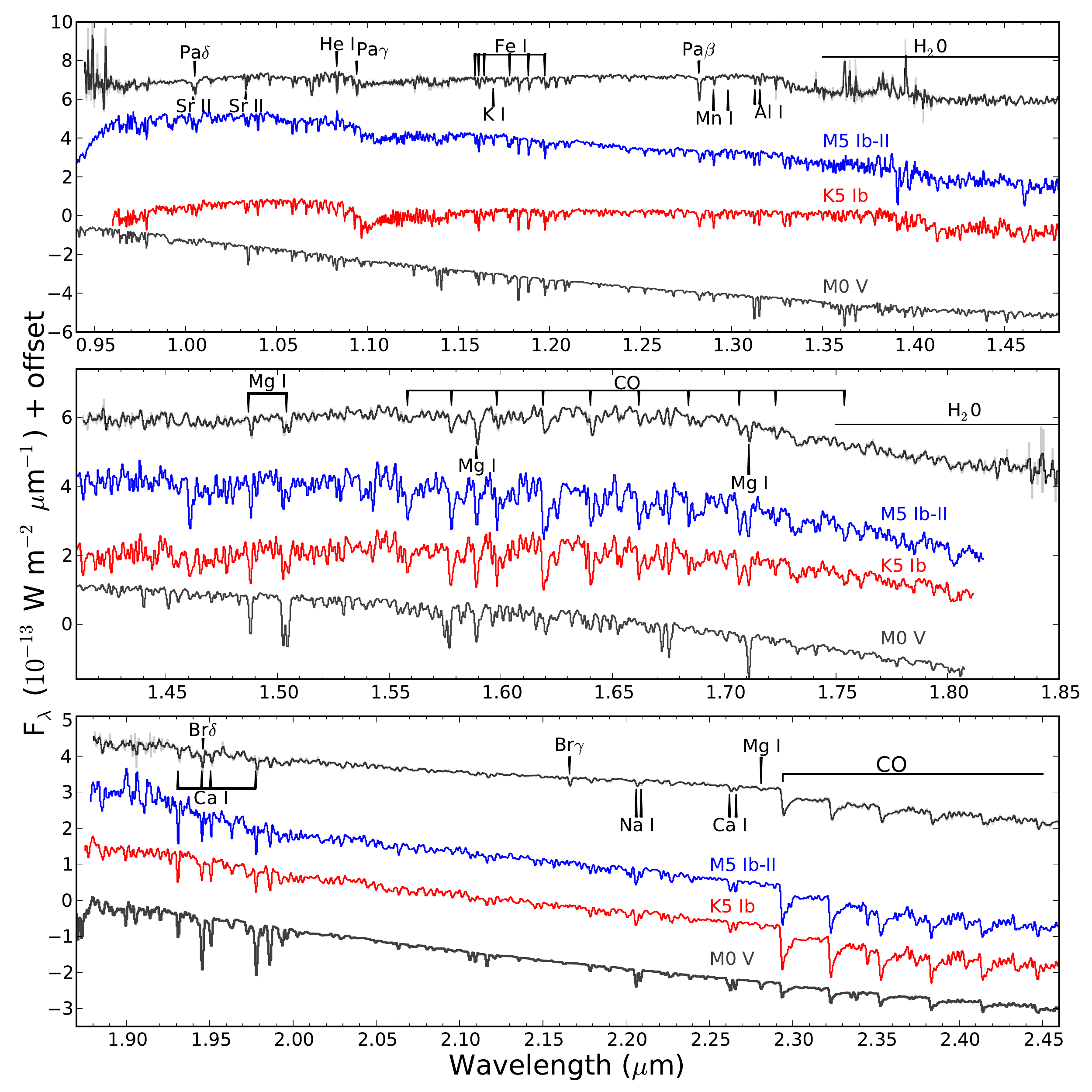}
\caption{TripleSpec NIR spectrum of \ptf\ showing from top to bottom
   the $YJ$, $H$, and $K$-band specta. Prominent absorption
  features in the outburst spectrum are identified, while the spectrum 
  shows no clear emission lines. For comparison the
  spectra of M5~Ib--II and K5~Ib supergiants, and an M0~V dwarf (from
  \citealt{rayner09}), are also shown. The muted features of \ptf\ 
  relative to the M0~V star shows that the NIR emission is from a low
  surface gravity environment. \ptf\ exhibits strong 2.3~$\mu$m CO
  absorption and $\sim$1.4 and 1.9 $\mu$m H$_2$O absorption, similar
  to other FU Orionis-like sources.  }
\label{NIR-spec}
\end{center}
\end{figure*}

The spectra were reduced with an IDL-based data reduction pipeline
developed by
P. Muirhead\footnote{http://www.astro.cornell.edu/$\sim$muirhead/\#Downloads.}. To
facilitate the subtraction of the sky and background emission of the
total signal, the observations were obtained using an ABBA dither
pattern along the slit. Each sky-subtracted exposure was then divided
by a normalized flat-field, wavelength calibrated, and optimally
extracted \citep{horne86}. The spectra were flux-calibrated and
corrected for telluric absorption using observations of an A0V star at
a similar airmass with the IDL-based code $xtellcor$\_$general$ by
\citet{vacca03}. The full TripleSpec spectrum is shown in
Figure~\ref{NIR-spec}.

\section{Analysis}\label{sec:analysis}

\subsection{Image Morphology and Photometric Analysis}\label{phot-analysis}

The large-amplitude eruption ($\sim$4 mag) of \lkh/\ptf\ qualitatively
matches the defining characteristic of FU Orionis eruptions. Such an
eruption, however, is not sufficient to prove that a YSO is in an FU
Orionis-like state. PTF 10nvg, for example, brightened by $\sim$6 mag
over the course of a year, yet shows little resemblance to FU Ori-like
stars (Covey et al. 2011).  Spectroscopy over a broad wavelength
range (see below) is required to conclusively classify a source as an
FU Orionis-like star.

Over the course of our P48 observations, \ptf\ brightened by $\sim$1
mag from early August 2009 to early August 2010, before suddenly
brightening by $\sim$3 mag over a $\sim$2 month period (see
Figure~\ref{P48-lc}). Between the start of our observations on 2009
Aug.\ 13 and 2010 Jul.\ 9, the initial slow rising phase, \ptf\ rose
by an average of $\sim$0.003 mag d$^{-1}$, while between 2010 Jul.\ 9
and 2010 Aug.\ 11, during the rapid rising phase, \ptf\ rose by an
average of $\sim$0.07 mag d$^{-1}$. 

After 2010 Aug.\ 11, \ptf\ is saturated in the P48 images and we
cannot constrain the rate at which it is increasing; however,
\citet{semkov10a} report that it reached peak brightness on 2010
Aug.\ 24 with $R = 12.79$ mag. Despite the rather rapid rise at the
beginning of the \ptf\ optical outburst, the overall light curve is
remarkably smooth. The P48 light curve for \ptf\ shows fluctuations at
the $\sim$0.2 mag level about both the pre-outburst slow rise and the
sharp rise of the outburst.  The long-term trend with low scatter
observed prior to 2010 July is somewhat unusual for a standard CTTS,
whose year-to-year variability is typically significantly smaller than
the intra-year variability \citep{grankin07}.  Indeed, the IPHAS
photometry of \lkh\ captures variability that is more characteristic
of CTTSs: variability at the $\sim$0.3 mag level was observed over the
course of 1 month in 2003, with a follow-up observation two years
later falling within the same range of magnitudes as observed in
2003. This suggests that the slow rise \lkh\ demonstrated in 2009 was
a prelude to the 2010 outburst, rather than simply standard CTTS
variability.

A new reflection nebula, which was not present in 2009, can now
clearly be seen around \ptf\ (Figure~\ref{reflect-nebula}). Our
regular observations over the course of the rise of \ptf\ allow us to
constrain the appearance of the nebula, similar to the analysis of
V1647 Ori/McNeil's Nebula by \citet{briceno04}. In the P48 subtraction
images produced as part of the PTF reduction pipeline, the residual is
 asymmetric about the point-source centroid on 2010
Aug.\ 2. The dust responsible for the reflection nebula is 
offset to the southwest of \ptf, and the asymmetry in the subtraction
image is the result of this dust scattering light into our line of
sight. We measure $R \approx 14$ mag for \ptf\ on 2010 Aug.\ 2. The
nebula may be present as early as 2010 July 19, however: the faint
nebula is difficult to distinguish from the wings of the central
star's point-spread function. The optical reflection nebula
(Figure~\ref{fig:P60_frame}) has an apparent size of $\sim$40\arcsec,
which at a distance of $\sim$600 pc corresponds to $\sim$24,000 AU.

In the NIR we have less information than in the optical. Our
observations suggest a flat light curve in the \jhk\ bands starting on
2010 Sep.\ 27.  Our NIR photometric measurements are consistent with
those presented by \citet{leoni10}, and we confirm that \ptf\ has
brightened by 3.25, 3.07, and 2.81 mag in the \jhk\ bands, respectively,
since 2MASS observations were obtained on 2000 Jun.\ 10. The NIR
measurements presented by \citet{cohen79} are $\sim$1.1 mag brighter
than those of 2MASS, which may be the result of a larger beam size for
the \citet{cohen79} measurements.  FU Orionis stars are frequently
associated with NIR reflection nebulae
\citep{connelley07,connelley10}, and \ptf\ is no exception (see
Figure~\ref{NIR-nebula}). The NIR nebula is similar in size to the
nebula seen in the optical.

The outburst SED is shown in Figure~\ref{progSED}, and it clearly
demonstrates that \ptf\ is bluer in the optical/NIR than it was prior
to the 2010 outburst. We have limited spectral coverage during the
outburst; in particular, we lack any information on how the emission
has changed in the MIR. Direct integration of the optical/NIR outburst
SED results in a lower limit to the luminosity of at least 4
\lsun. Assuming the same bolometric correction to the $H$ band before
and after the outburst results in an outburst luminosity of $L_{\rm
  outburst} \approx 12$ \lsun. This luminosity places \ptf\ at the
bottom of the luminosity scale for FU Ori outbursts, and more than an
order of magnitude fainter than FU Ori itself \citep{hartmann96}.

\subsection{Spectroscopic Analysis}

Optical spectra of FU Orionis variables share a few defining
properties: a late F or G supergiant photosphere, an
absorption-dominated spectrum with only H$\alpha$ typically seen in
emission, and broad P Cygni absorption with representative velocities
of a few hundred \kms. Figure~\ref{spectrum} shows that
\ptf\ exhibits these general properties. Figure~\ref{HIRES-lines}
illustrates our HIRES spectra covering three Balmer transitions. Each
of these lines exhibits a strong P Cygni absorption profile, with
absorption minima around $-40$ \kms.  H$\alpha$ is the only line
having an emission component, and this emission moves the absorption
minimum slightly blueward of the other Balmer lines. The absorption
extends to $\la-$200 \kms, suggesting in the classical interpretation
that \ptf\ is driving a strong outflow (see also
Figure~\ref{HIRES-lines}).

In the discussion below, all velocities have been transformed to the
local standard of rest (LSR). To our knowledge, a precise radial
velocity measurement for \lkh\ has not been made. Therefore, we adopt
a rest velocity of $V_{\rm LSR} \approx 1.6$ \kms, consistent with
$^{13}$CO observations of the surrounding molecular gas
\citep{dobashi94}.

\subsubsection{Low-Resolution Optical Spectroscopy}

At low resolution, \ptf\ exhibits little evolution between our two
Kast spectra taken 40~d apart (see Figure~\ref{spectrum}). The
strength of the absorption features remains constant to within the
uncertainties, while the only major change is in the emission flux
from H$\alpha$.  During this 40~d period the H$\alpha$ flux has
decreased from $\sim 2.8 \times 10^{30}$ erg s$^{-1}$ to $\sim 1.5
\times 10^{30}$ erg s$^{-1}$. Observing at the parallactic angle
allows us to accurately measure the change in the relative color
during the 40~d gap between observations. Given the typical
uncertainties associated with spectrophotometric measurements with the
Kast spectrograph (Silverman et al., 2011, in prep.), we find that
\ptf\ does not show significant evidence for a change in color between
our two observations.

To constrain the optical spectral type of \ptf, we use a least-squares
minimization procedure to find the best match between the
low-resolution Kast spectrum and the low-resolution stellar spectra
provided in the Pickles Stellar Spectral Flux Library
\citep{pickles98}. The spectral type of FU Orionis variables is known
to vary as a function of wavelength \citep{herbig77}; thus, we divide
the spectra into two sections, 4000--6000~\AA\ and
6000--9000~\AA. Below 4000~\AA, the $S/N$ is low in our Kast spectrum,
while above $\sim$9000~\AA\ there is an uncertain telluric correction
which may add artificial features. Next we redden each of the Pickles
star spectra in increments of 0.05 mag from \ebv\ $=$ 0.00 to 2.25 mag
with $R_V = 3.1$. We mask the region around H$\alpha$, since emission
lines are not expected in normal stellar spectra. We then perform a
least-squares fit of the reddened library spectra with our
low-resolution spectrum of \lkh, where the only free parameter is an
overall normalization constant.

\begin{figure*}
\begin{center}
%\epsscale{0.9}
\includegraphics[width=145mm]{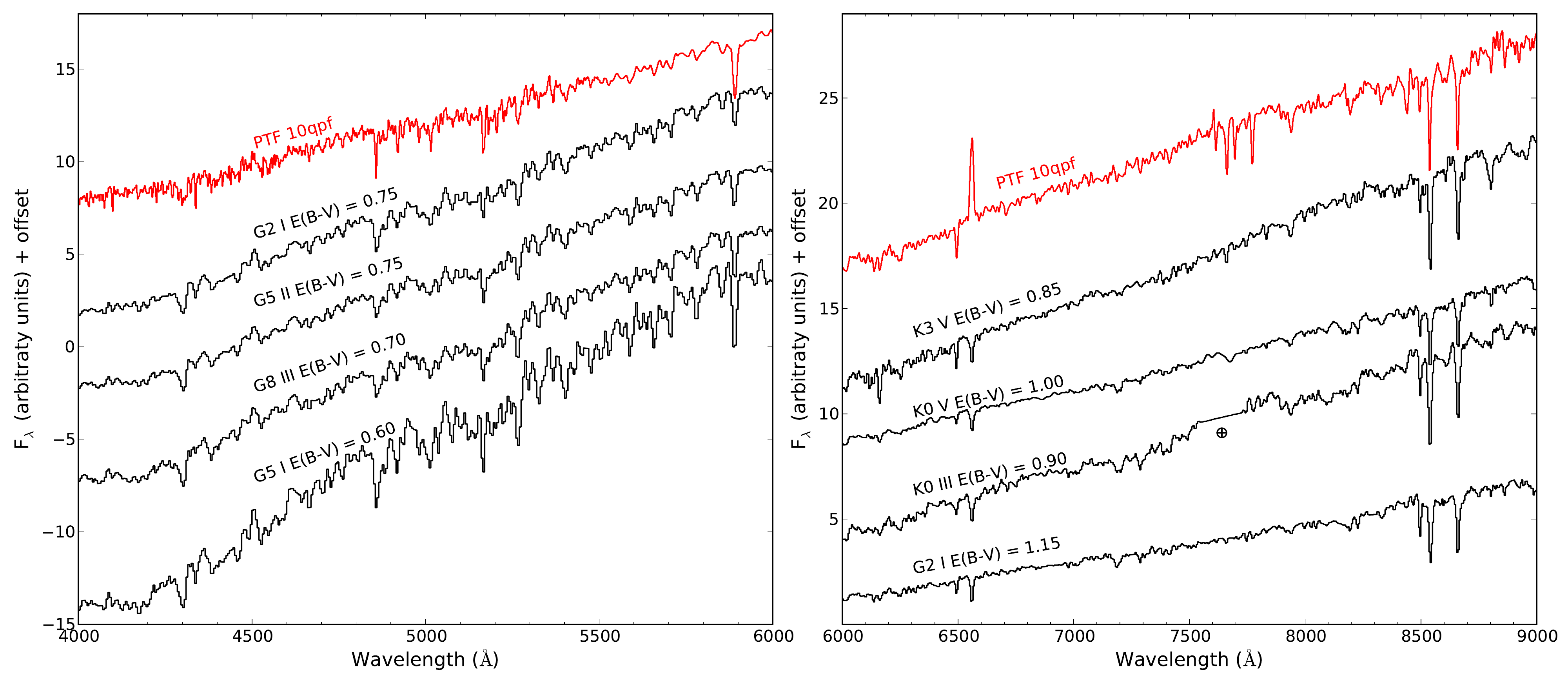}
\caption{Low-resolution optical spectra of \ptf\ compared to stellar
  spectra from the Pickles library \citep{pickles98}. The comparison
  spectra are selected as the best matches to \ptf\ following a
  least-squares minimization procedure (see text). The comparison
  spectra are shown in order of the goodness of fit, with the best
  match on top. The reddening values were used as a proxy for removing 
  the overall continuum shape.  }
\label{opt-low-res}
\end{center}
\end{figure*}

We illustrate the results of this procedure in
Figure~\ref{opt-low-res}, where we include the low-resolution spectrum
of \ptf\ along with the four best matches from our least-squares
minimization. The stellar spectra are shown in the order of the
quality of the fit, with the best fit on top and the fourth-best fit
on the bottom. The adopted reddening has little physical meaning given
that the observed flux is not emanating from a stellar photosphere
having the standard single-temperature SED; this portion of the
procedure merely serves as an automated process to vary the slope of
the continuum, which is necessary given that \ptf\ has a heavily
extinguished spectrum.

The Pickles library is incomplete --- it does not include every
luminosity class for every spectral type. For this reason we emphasize
that our procedure is only illustrative; it does not provide a
definitive spectral type for \ptf\ in either the blue or red portion
of the optical spectrum. Nevertheless, a few trends are readily
apparent from Figure~\ref{opt-low-res}. In the blue portion of the
optical, \ptf\ most closely resembles the spectra of G supergiants and
G giants.  In the red portion of the optical, the spectrum resembles
that of G and K stars, with no real hint regarding the luminosity
class. The adopted values of the reddening are significantly different
between the blue and the red, but again we remind the reader that this
reddening is only a device with which to match the intrinsic shape of
the continuum. Typically, the Mg~I $\lambda$5172 triplet is used to
determine the luminosity class of G-type stars, but our
high-resolution spectra show that these lines are tracing an outflow
(see Figure~\ref{HIRES-lines}) and thus they are unsuitable for
providing a luminosity class. Despite this, the procedure does
demonstrate that \ptf\ is similar to G giants/supergiants in the blue
portion of the optical, consistent with observations of other FU
Orionis variables \citep{hartmann96}. Additionally, as we will show
below, the spectrum shifts to later types (K stars in this case) with
increasing wavelength, as is also observed in FU Orionis variables
\citep{herbig77}.

\subsubsection{High-Resolution Optical Spectra}

The HIRES spectrum of \ptf\ is dominated by broadened stellar-like
absorption features. The spectral lines are consistent with those of a
comparison G2~I star and inconsistent with spectral types later than
early K. We note, however, that the diagnostics of G vs. K types are
subtle and would become blurred with the large broadening of the
\ptf\ lines.  The line broadening is similar to, but perhaps slightly
less than, the 60--100 \kms\ broadening of V1057 Cyg, and
significantly more than the 30--40 \kms\ of V1515 Cyg in which some
hints of a K-type spectrum can be seen at redder optical wavelengths.
A detailed comparison of \ptf\ to these two canonical FU Ori objects
reveals that the spectrum of \ptf\ is similar to both in terms of the
relative strength of the common absorption features. However, we note
that each object has unique and notable spectroscopic attributes.

\begin{figure*}
\begin{center}
%\epsscale{0.9}
\includegraphics[width=160mm]{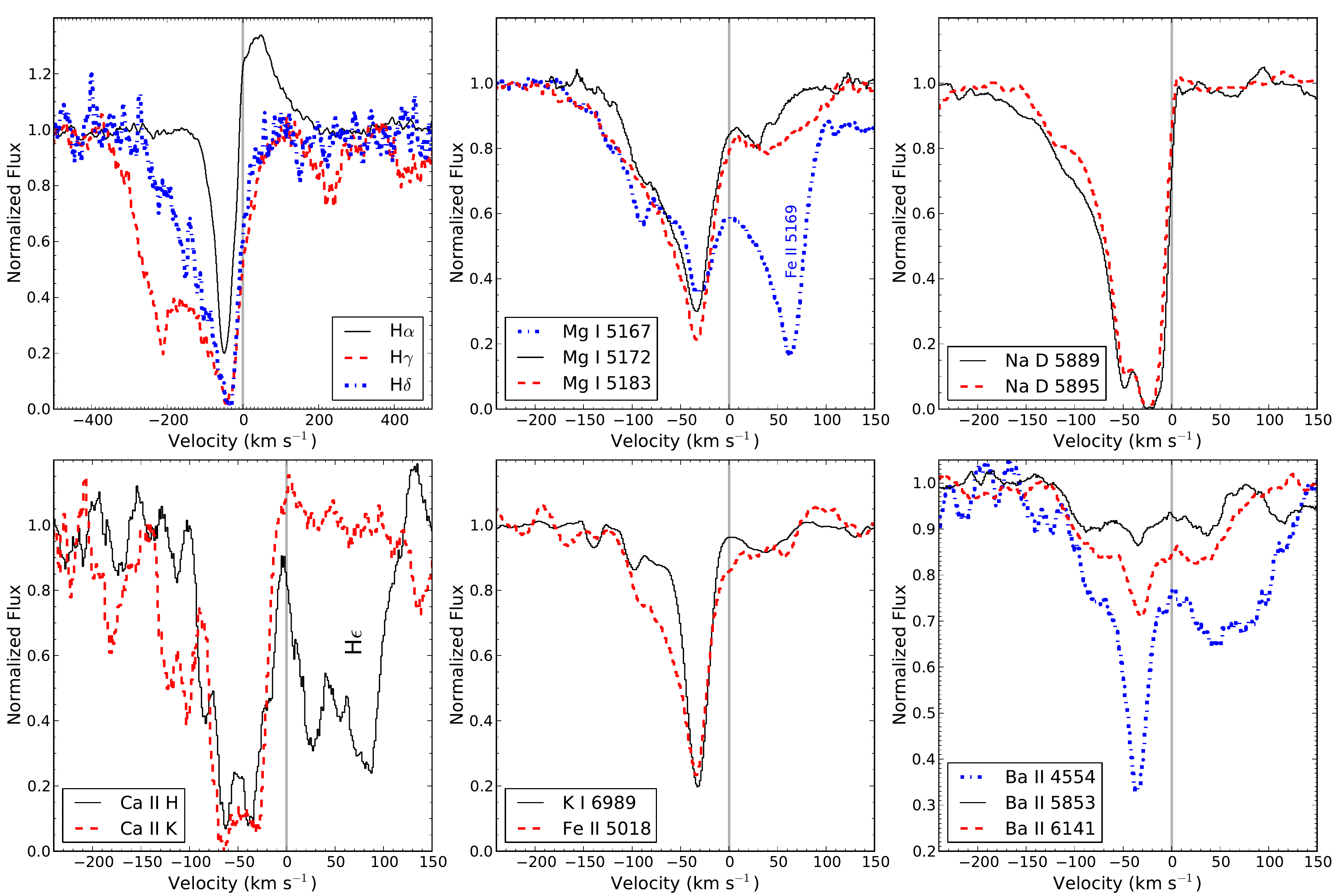}
\caption{Keck HIRES high-resolution profiles of several prominent
  features showing strong blueshifted absorption, indicative of
  outflowing material.  {\it Top left}: The Balmer series for
  \ptf. H$\beta$ fell in a gap between echelle orders for our
  observing setup, and thus is not displayed. \ptf\ exhibits strong P
  Cygni absorption in the other Balmer lines with outflow speeds of
  $\sim$100--300 \kms.  {\it Top center}: Profiles of the Mg~I triplet
  at 5167, 5172, and 5183~\AA. The lines show strong asymmetries,
  tracing the kinematics of the wind from \ptf.  Fe~II $\lambda$5169
  is present at about +60 \kms\ relative to Mg~I $\lambda$5167. {\it
    Top right}: Absorption profiles of the Na~I~D doublet
  showing similar kinematics as the Mg I lines. {\it Bottom
    left}: Absorption profiles of Ca~II H \& K. H$\epsilon$ is also
  present, offset from Ca~H by $\sim$100 \kms. {\it Bottom center}:
  Absorption profiles of Fe~II $\lambda$5018 and K~I
  $\lambda$6989. {\it Bottom right}: Absorption profiles of Ba~II
  $\lambda\lambda$4554, 5853, and 6141. These lines show narrow
  absorption, FWHM $\approx$ 20 \kms, superposed on broader, shallower 
  profiles extending $\pm\sim$60 \kms\ from the line center. For
  clarity, all spectra have been smoothed with a Savitzky-Golay
  filter, which is similar to a running mean
  \citep{savitzky64}. Velocities are given relative to the LSR, and
  the vertical grey line indicates the rest velocity of \ptf, $V_{\rm
    LSR} \approx 1.6$ \kms\ (see text).  }
\label{HIRES-lines}
\end{center}
\end{figure*}

Several lines have clear kinematic signatures of winds, specifically
Ca~II H \& K, Fe~II $\lambda$5018, the Mg~I $\lambda\lambda$5167,
5172, 5183 triplet, the Na~I~D lines, H$\alpha$, H$\gamma$, H$\delta$
(H$\beta$ falls in a gap between the orders in our HIRES setting), and
the K~I $\lambda$7699 line (its doublet partner at 7665~\AA\ also
falls in a gap between orders).  The morphology of these lines is
illustrated in Figure~\ref{HIRES-lines}.  Notably, the blueshifted
absorption troughs are narrower in our early-stage spectrum of
\ptf\ than exhibited in modern spectra of V1515 Cyg and V1057 Cyg, by
at least a factor of two.  Although \lkh\ exhibited [O~I]
$\lambda$6300 emission in the \citet{cohen79} spectrum, it is not
apparent post-outburst, even at high spectral resolution, nor do we
see other forbidden species in emission (such as [N~II] and [S~II])
typically associated with outflows in YSOs.

\begin{figure}
\begin{center}
%\epsscale{0.9}
\includegraphics[width=85mm]{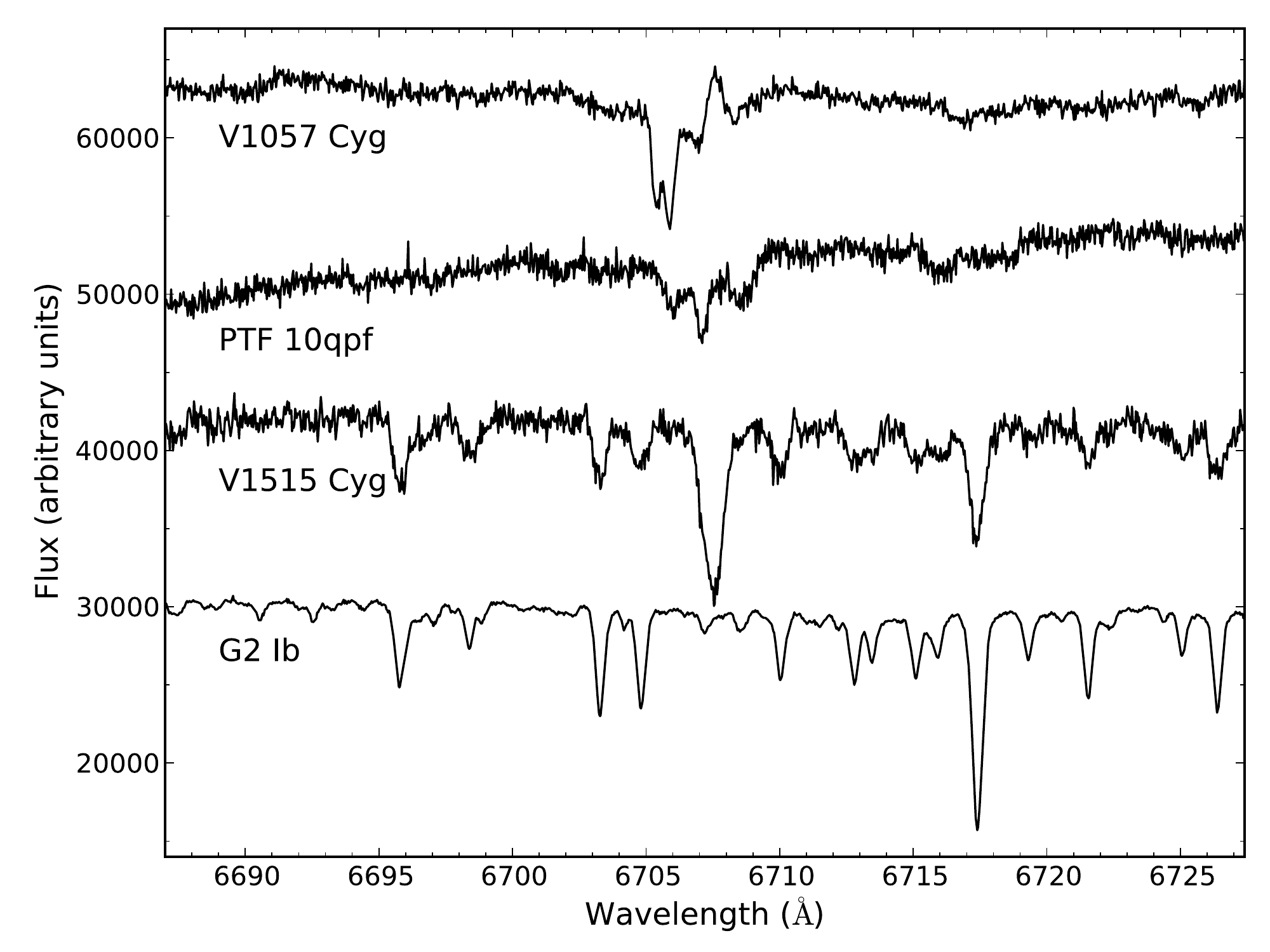}
\caption{Keck HIRES high-resolution spectra of the Li~I $\lambda$6706
  and Ca~I $\lambda$6717 region of \ptf.  For comparison we also show
  high-resolution spectra of V1057 Cyg, V1515 Cyg, and a G2~Ib
  spectral standard, all from our unpublished spectral database.  In
  particular, notice the strong Li absorption present in \ptf\ and the
  two canonical FU Orionis objects. The Li line in both V1057 Cyg and
  \ptf\ traces an unusual kinematic structure, possibly similar to
  that seen in Ba~II (see Figure~\ref{HIRES-lines}.)  }
\label{LiSpec}
\end{center}
\end{figure}

The other line worth noting in \ptf\ for its kinematic signatures is
Li~I $\lambda$6707.  Figure~\ref{LiSpec} shows the spectral region
containing this line, in comparison to V1515 Cyg and V1057 Cyg as well
the G2~I standard.  The Li I absorption profile has multiple
components including a narrow slightly blueshifted component 
arising from low-velocity
gas and a broader absorption extending to $\pm 60$ \kms\ that is
reminiscent of the flat-bottomed profiles seen in other metallic lines
in the same spectral region. A similar absorption profile is also seen
in Ba~II (Figure~\ref{HIRES-lines}).

\begin{figure}
\begin{center}
%\epsscale{0.9}
\includegraphics[width=85mm]{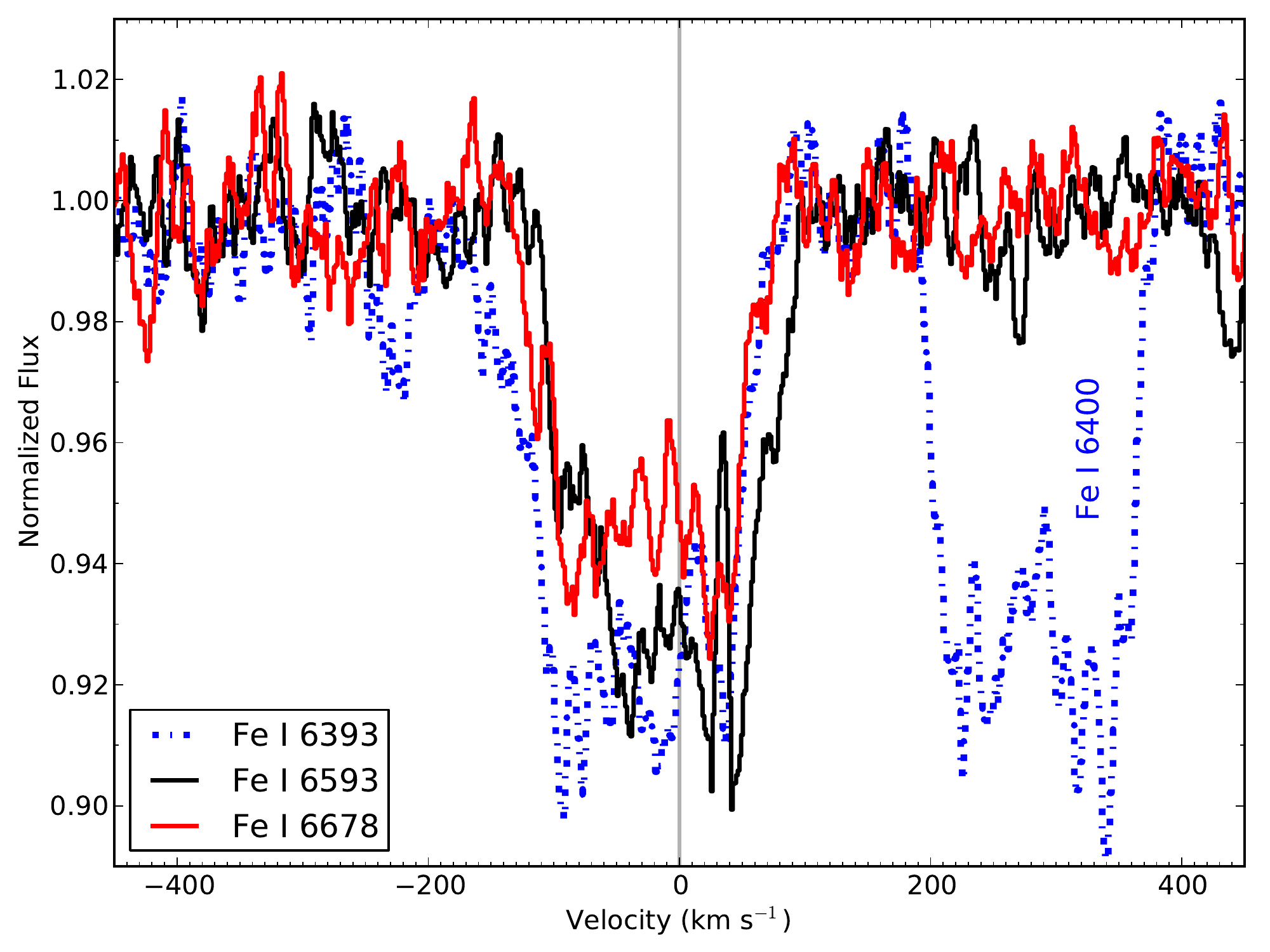}
\caption{Several Fe~I line profiles illustrating the flat-bottomed
  nature typical of the other metallic lines. The Fe~I transitions at
  $\lambda\lambda$6393, 6593, and 6678 have been shifted to the same
  velocity, while Fe~I $\lambda$6400 is offset by $\sim$300
  \kms\ relative to $\lambda$6393. These lines all exhibit a profile
  with a ``flat'' bottom similar to those observed in FU Orionis
  \citep{petrov08}. For clarity, spectra have been smoothed with a
  Savitzky-Golay filter, which is similar to a running mean. The
  vertical grey line is the same as in Figure~\ref{HIRES-lines}.}
\label{FeSpec}
\end{center}
\end{figure}

Figure~\ref{FeSpec} illustrates the flat-bottomed nature of many
absorption lines in \ptf.  The profiles are reminiscent of those
illustrated by \citet{petrov08} for FU Ori.  When the flat-bottomed
lines in \ptf\ are cross-correlated against a normal G2~I stellar
template, the resulting cross-correlation function is broad ($\sim$120
\kms) with several different peaks. In some spectral regions a central
peak can be fit by a Gaussian of width 30 \kms, and there are two
secondary peaks near the full width of the correlation function, each
having Gaussian width $\sim$20 \kms. In other spectral regions, both
blue and red, the correlation function has no central peak, but only
the double peaks near the limb. To measure the goodness of the cross
correlation we adopt the Tonry-Davis $R$ value \citep{tonry79}, which
is a ratio of the power in the cross-correlation peak to a measure of
the antisymmetry in the correlation. Large $R$ values indicate good
cross correlations. Using a G2 supergiant as a template, the
Tonry-Davis $R$ value for \ptf\ is 4--6 (compared to $R > 5$ to 15 for
V1057 Cyg, which shows broadening and flat-bottomed structure similar
to \ptf\ in some lines, and $R > 20$ for V1515 Cyg, which does
not). The cross-correlation results are similar when using V1515 Cyg
as a template, rather than the G2~I standard. Employing V1057 Cyg as a
template results in a broad and centrally peaked cross-correlation
function having width 120--130 \kms.

\subsubsection{Near-Infrared Spectroscopy}

NIR spectra of FU Orionis variables resemble the spectra of K and M
giants/supergiants, in contrast to their earlier spectral type in the
optical \citep{hartmann96}. The most prominent features in the NIR
spectra of FU Orionis stars are the strong absorption bands from CO at
2.3 $\mu$m and the H$_2$O vibration-rotation bands at 1.38 $\mu$m and
1.87 $\mu$m. These absorption bands are clearly present in the
outburst spectrum of \ptf; see Figure~\ref{NIR-spec}, which compares
spectra from \citet{rayner09} of two supergiants (K5~Ib and M5~Ib-II)
and a dwarf (M0~V) with \ptf. There are many prominent absorption
features seen in the M0~V spectrum, which are muted
or absent in \ptf\ and the two supergiant spectra. This is indicative 
of a low surface gravity environment for \ptf.

We have attempted to derive a crude NIR spectral type for \ptf.  In
addition to its anomalously strong CO features, \ptf\ also possesses
prominent H$_2$O absorption in the $H$ and $K$ bands; the relative
strengths of the $H$ and $K$-band absorptions are unusual, with
\ptf\ separated from both the dwarf and giant loci for those features
(see, e.g., Figure 4 of \citealt{covey10}).  We therefore attempted to
characterize the NIR spectral type of \ptf\ based on the strengths of
several $H$-band absorption features: Mg~I (1.51 $\mu$m), K~I (1.518
$\mu$m), Al~I (1.67 $\mu$m), and a blend at 1.70 $\mu$m. We compared
the strengths of these features to those measured for a suite of dwarf
and giant templates by \citet{covey10}.  While this method is
sufficient to derive only a crude NIR spectral type for \ptf,
we find that the source's outburst spectrum most closely resembles
that of K7~III---M2~III giants (see Figure~\ref{NIR-spec} for some
comparisons). It is important to note that the classification scheme
from \citet{covey10} does not include supergiants; thus, we conclude
that \ptf\ has a low surface gravity, but we cannot discriminate among
luminosity classes I--III.

While the strength of molecular water absorption is greater than that
normally observed in giants, this finding is consistent with
observations of other FU Orionis stars. \citet{sato92} argue that the
unusually strong H$_2$O absorption seen in FU Orionis stars can be
explained if water that has condensed onto dust grains in the midplane
of the disk becomes vaporized during the heating of the disk
associated with an FU Orionis eruption. Once this evaporated water is
flung vertically out of the disk midplane, it will imprint a strong
absorption signature in the NIR spectrum.

\citet{connelley10} show that the equivalent width (EW) of CO
absorption at \hbox{2.3 $\mu$m} can be used to identify FU
Orionis-like stars when compared to the EW of Na (2.206 and
\hbox{2.209 $\mu$m}) and Ca (2.263 and 2.266 $\mu$m). To compare
\ptf\ to the stars studied by \citet{connelley10}, we measure the EW
of Na, Ca, and CO following the prescription of \citet{chalabaev83}
with the correction provided by \citet{vollmann06}. We measure Na and
Ca EWs of 1.56 $\pm$ 0.05 and 1.07 $\pm$ 0.06 \AA, respectively, while
the EW of CO is 35.6 $\pm$ 0.3 \AA. These uncertainties reflect the
statistical errors in our measurements, and likely underestimate the
systematic effects associated with these measurements (e.g., the
precise placement of the continuum).  Nevertheless, the systematic
uncertainties are not large enough to alter the general conclusions
drawn below. The CO absorption is well in excess of what would be
expected for both late-type dwarfs and giants based on the strength of
the Na and Ca absorption; however, this excess CO absorption is
consistent with other FU Orionis-like stars in the sample of Connelley
\& Greene (see their Figure~4).

Prominent H and He I $\lambda$10830 absorption lines are also visible
in the NIR spectra (see Figure~\ref{NIR-spec}).  However, H and He
absorption cannot originate in a simple, single-temperature, cool (late
K--M) atmosphere. Thus, these lines, though unresolved in our spectra,
likely trace an outflow, similar to the Balmer series in the optical.
He I $\lambda$10830 may be a hot-wind indicator, but high-resolution
spectra are needed to confirm this \citep{edwards06,kwan07}.

\section{Discussion and Conclusions}\label{sec:conclusions}

We have shown that the 2010 eruption of \lkh\ places 
it within the small family of FU Orionis-like stars having observed
outbursts. \lkh/\ptf\ exhibits the defining observable
characteristics of the FU Orionis class as described by
\citet{hartmann96}:
\begin{itemize}
	\item it exhibited a $\ga$ 4 mag increase in its optical light
          curve;
	\item it is associated with a star-forming region and shows a
          bright reflection nebula following its eruption;
	\item optical spectra of the outburst are consistent with a
          G-type giant/supergiant;
	\item during outburst, the Balmer lines exhibit strong P Cygni
          profiles, with only H$\alpha$ showing an emission component,
          and absorption extending to about $-$200 \kms; and
	\item the NIR spectra during the eruption resemble those of
          M-type giants/supergiants, with strong absorption seen in
          the H$_2$O and CO bands.
\end{itemize}
In sharp contrast to the class of YSO outbursts referred to as EX Lupi-like
or V1647 Ori-like, \ptf\ does not exhibit strong optical/NIR emission
lines during outburst.

In the IR, unlike typical M and K-type supergiants, \ptf\ shows
prominent absorption lines of H I (Br$\gamma$, Pa$\beta$, Pa$\gamma$,
Pa$\delta$) and He I 1.083 $\mu$m. In the optical the H I lines trace
an outflow, with clear P Cygni absorption profiles observed in our
HIRES spectrum. Our spectral resolution is insufficient to determine
whether a similar P Cygni profile is observed in the NIR; however, H
and He absorption cannot originate in an M-giant/supergiant
photosphere. Thus, the H and He lines are likely the result of 
inner-disk region accretion/wind physics.

\lkh\ stands out among the currently known group of FU Orionis-like stars
in that there are pre-outburst observations with detections from the
optical to the MIR, making \lkh\ the first FU Orionis-like star with a
well-sampled SED prior to eruption. This SED shows that \lkh\ was a
Class II YSO, which are commonly associated with CTTSs
\citep{lada87}. A pre-outburst optical spectrum reveals a late-type
star, K7--M0, with prominent H$\alpha$ and H$\beta$ seen in emission,
again consistent with a CTTS. Multiple epochs of optical imaging prior
to the outburst reveal that \lkh\ exhibited stochastic variability at
a level consistent with a CTTS.

One of the most important elements in the study of \ptf/\lkh\ is that it is
associated with a known, optically revealed and previously studied
YSO.  The clear identification of \lkh\ as a CTTS prior to eruption
solidifies the interpretation of FU Ori events as enhanced accretion
and outflow likely associated with disk-accretion instabilities
\citep{hartmann96}. This outburst's association with a previously
classified CTTS source shows that FU Ori eruptions are not strictly
limited to the most highly embedded Class 0/I phases of pre-main
sequence evolution. Indeed, the strong winds associated with
these events may play an important role in disrupting a YSO's
circumstellar envelope. Such a disruption is thought to be a key
component of the Class I/II transition.

Continued photometric and spectroscopic monitoring of \ptf/\lkh\ will 
fully elucidate its nature. As more and more synoptic surveys come 
online, the rich legacy of IR
observations made by several ground- and space-based observatories
(e.g., 2MASS, {\it IRAS}, {\it MSX}, {\it Spitzer}, {\it AKARI}, {\it
  WISE}) over the past two decades will enable the discovery of
additional FU Orionis outbursts with well characterized pre-outburst
SEDs. These discoveries will in turn allow us to determine if the case
of \lkh\ is an outlier or the norm. A better understanding of whether
FU Orionis-like eruptions typically occur during the Class I or Class II
phase of pre-main sequence evolution will allow us to place stronger
constraints on the triggering mechanism and occurrence rates for these
events.
 
\acknowledgments 

We thank Luisa Rebull for providing the {\it Spitzer}/MIPS data for 
\lkh\ in advance of her publication. 
We are grateful to Michael Kandrashoff and Alekzandir Morton for their 
assistance in obtaining Kast spectra. We are in debt to John Johnson, 
John Sebastian Pineda, and Michael Bottom who obtained the HIRES spectrum 
for us. We thank Cullen Blake, Dan Starr, and Emilo Falco for their 
efforts to build and maintain PAIRITEL. We would like to thank Meredith 
Hughes for a fruitful discussion concerning YSO accretion disks and 
Ryan Foley for information on the spectrophotometric uncertainty 
associated with Kast spectra. An anonymous referee provided comments 
that have improved this manuscript.

A.A.M. is supported by the National Science Foundation (NSF) Graduate
Research Fellowship Program. K.R.C. acknowledges support for this work
from the Hubble Fellowship Program, provided by the National
Aeronautics and Space Administration (NASA) through Hubble Fellowship
grant HST-HF-51253.01-A awarded by the STScI, which is operated by the
AURA, Inc., for NASA, under contract NAS 5-26555. J.S.B. acknowledges
support of an NSF-CDI grant-0941742. A.A.M. and C.R.K. were partially
supported by NSF-AAG grant-1009991. A.V.F.'s group is grateful for the
support of NSF grant AST-0908886, the TABASGO Foundation, Gary and
Cynthia Bengier, and the Richard and Rhoda Goldman Fund. The National 
Energy Research Scientific Computing Center, which is supported by the 
Office of Science of the U.S. Department of Energy under Contract No. 
DE-AC02-05CH11231, provided staff, computational resources, and data 
storage for this project.

Some of the data presented herein were obtained at the W. M. Keck
Observatory, which is operated as a scientific partnership among the
California Institute of Technology, the University of California, and
NASA. The Observatory was made possible by the generous financial
support of the W. M. Keck Foundation. The authors wish to recognize
and acknowledge the very significant cultural role and reverence that
the summit of Mauna Kea has always had within the indigenous Hawaiian
community.  We are most fortunate to have the opportunity to conduct
observations from this mountain.

PAIRITEL is operated by the Smithsonian Astrophysical Observatory
(SAO) and was made possible by a grant from the Harvard University
Milton Fund, a camera loan from the University of Virginia, and
continued support of the SAO and UC Berkeley.  The PAIRITEL project
and those working on PAIRITEL data are further supported by NASA/Swift
Guest Investigator Programs NNX09AQ66Q and NNX10A128G. We are grateful
for the assistance of the staffs at all of the observatories used to
obtain the data.

This research has made use of NASA's Astrophysics Data System
Bibliographic Services, the SIMBAD database operated at CDS,
Strasbourg, France, the NASA/IPAC Extragalactic Database operated by
the Jet Propulsion Laboratory, California Institute of Technology,
under contract with NASA, and the VizieR database of astronomical
catalogs \citep{Ochsenbein2000}. This publication makes use of data
products from the Two Micron All Sky Survey, which is a joint project
of the University of Massachusetts and the Infrared Processing and
Analysis Center/California Institute of Technology, funded by NASA and
the NSF.

Facilities:
\facility{Palomar: 48 inch (PTF);}
\facility{Palomar: 60 inch (PTF);}
\facility{Palomar: 200 inch (TripleSpec);}
\facility{Keck (HIRES);}
\facility{Lick: 3 m (Kast);}
\facility{FLWO: 1.3 m (PAIRITEL).}

%\bibliographystyle{apj1c}

%\bibliography{/Users/amiller/Desktop/papers}

\newpage

\begin{table}
\begin{minipage}{80mm}
  \centering
\caption{IPHAS Observations of \lkh. 
}
\begin{tabular}{rccc}
\hline
UT date & $r$ mag & $i$ mag & H$\alpha$ mag \\ 
   & (AB) & (AB) & (AB) \\
\hline
2003/10/15 & 17.300$\pm$0.010 & 15.677$\pm$0.006 & 16.023$\pm$0.007 \\
2003/10/18 & 17.348$\pm$0.011 & 15.744$\pm$0.006 & 16.063$\pm$0.006 \\
2003/11/02 & 17.283$\pm$0.017 & 15.608$\pm$0.009 & 15.975$\pm$0.009 \\
2003/11/14 & 17.074$\pm$0.009 & 15.570$\pm$0.005 & 15.724$\pm$0.006 \\
2005/11/19 & 17.075$\pm$0.007 & 15.671$\pm$0.006 & 15.768$\pm$0.008 \\	
\hline
\end{tabular}
\label{tab-iphas}
%\end{center}
\end{minipage}
\end{table}

% \begin{deluxetable}{lcc}
% %\tablewidth{0pt}
% % \tabletypesize{\tiny}
% \tablecaption{P48 $R$-Band Photometry of \ptf \label{tab-P48}}
% \tablehead{
% \colhead{date}  & 
% \colhead{mag} & 
% \colhead{$\sigma_{\rm mag}$} \\
% \colhead{(MJD)}  & 
% \colhead{} & 
% \colhead{} 
% }
% \startdata

\begin{table}
\begin{minipage}{80mm}
  \centering
\caption{P60 Observations of \ptf. 
}
\begin{tabular}{rccc}
\hline
date & filter & mag & $\sigma_{\rm mag}$ \\ 
(MJD) &  & (AB) & \\
\hline
55483.127 & $r$ & 12.948 & 0.039 \\
55483.127 & $i$ & 12.207 & 0.027 \\
55483.128 & $z$ & 11.623 & 0.041 \\
\hline
\end{tabular}
\label{tab-P60}
%\end{center}
\end{minipage}
\end{table}

\begin{table}
\begin{minipage}{80mm}
  \centering
\caption{PAIRITEL Observations of \ptf. 
}
\begin{tabular}{rccc}
\hline
$t_{\rm mid}$\footnote{Midpoint between the first and last exposures in a single stacked image.} & $J$ mag & $H$ mag & $K_s$ mag \\ 
(MJD) & (Vega) & (Vega) & (Vega) \\
\hline
55466.137 & 10.04 $\pm$ 0.03 &  9.14 $\pm$ 0.03 &  8.65 $\pm$ 0.03 \\
55468.145 &  9.99 $\pm$ 0.03 &  9.06 $\pm$ 0.04 &  8.64 $\pm$ 0.04 \\
55469.148 & 10.04 $\pm$ 0.03 &  9.07 $\pm$ 0.03 &  8.70 $\pm$ 0.03 \\
55479.109 & 10.11 $\pm$ 0.03 &  9.11 $\pm$ 0.03 &  8.63 $\pm$ 0.04 \\
55504.164 & 10.20 $\pm$ 0.03 &  9.24 $\pm$ 0.03 &  8.74 $\pm$ 0.03 \\
55513.195 & 10.29 $\pm$ 0.03 &  9.34 $\pm$ 0.03 &  8.79 $\pm$ 0.03 \\
55518.168 & 10.32 $\pm$ 0.03 &  9.34 $\pm$ 0.04 &  8.84 $\pm$ 0.03 \\
55527.117 & 10.35 $\pm$ 0.03 &  9.40 $\pm$ 0.03 &  8.89 $\pm$ 0.03 \\
55531.145 & 10.40 $\pm$ 0.03 &  9.44 $\pm$ 0.03 &  8.97 $\pm$ 0.03 \\
55543.066 & 10.45 $\pm$ 0.03 &  9.48 $\pm$ 0.04 &  9.06 $\pm$ 0.04 \\
\hline
\end{tabular}
\label{tab-ptel}
%\end{center}
\end{minipage}
\end{table}

\begin{table}
\begin{minipage}{100mm}
  \centering
\caption{Log of spectroscopic observations.}
\begin{tabular}{cccc}
\hline
UT Date & Telescope/ & Exposure & $\lambda$ \\
 & Instrument & (s)  & (\AA)\\
\hline
2010-09-16.24 & Shane 3-m/Kast & 600 & 3400--10300 \\
2010-09-23.24 & Hale 5-m/TripleSpec & 720 & 9400--24,600 \\
2010-09-25.30 & Keck I 10-m/HIRES & 560 & 3640--7990 \\
2010-11-02.21 & Shane 3-m/Kast & 1200 & 3400--10300 \\
\hline
\end{tabular}
\label{speclog}
\end{minipage}
\end{table}

\begin{table}
\begin{minipage}{80mm}
  \centering
\caption{P48 $R$-Band Photometry of \ptf
}
{\tiny
\begin{tabular}{lcc}
\hline
{\footnotesize date} & {\footnotesize mag} & {\footnotesize $\sigma_{\rm mag}$} \\
{\footnotesize (MJD)} & & \\
\hline
55056.221 & 17.023 & 0.017 \\
55056.295 & 17.098 & 0.019 \\
55061.383 & 16.843 & 0.012 \\
55062.327 & 16.880 & 0.020 \\
55064.286 & 16.950 & 0.018 \\
55064.331 & 16.818 & 0.021 \\
55067.393 & 16.803 & 0.031 \\
55080.361 & 17.079 & 0.036 \\
55093.318 & 16.933 & 0.016 \\
55094.252 & 16.777 & 0.018 \\
55107.222 & 16.981 & 0.026 \\
55107.266 & 16.963 & 0.025 \\
55123.225 & 16.942 & 0.013 \\
55302.479 & 16.424 & 0.010 \\
55303.411 & 16.247 & 0.012 \\
55303.459 & 16.198 & 0.014 \\
55310.370 & 16.053 & 0.022 \\
55310.413 & 16.164 & 0.010 \\
55316.420 & 16.372 & 0.018 \\
55322.359 & 15.714 & 0.017 \\
55322.403 & 16.308 & 0.009 \\
55328.453 & 15.988 & 0.011 \\
55335.477 & 16.124 & 0.010 \\
55340.429 & 16.140 & 0.008 \\
55340.474 & 16.102 & 0.007 \\
55345.470 & 16.182 & 0.016 \\
55346.270 & 16.058 & 0.015 \\
55346.314 & 16.157 & 0.017 \\
55351.271 & 16.196 & 0.010 \\
55356.309 & 16.165 & 0.013 \\
55356.353 & 16.141 & 0.008 \\
55361.311 & 16.156 & 0.009 \\
55361.355 & 16.163 & 0.011 \\
55366.318 & 16.046 & 0.012 \\
55366.362 & 16.026 & 0.009 \\
55371.384 & 15.963 & 0.011 \\
55376.377 & 16.119 & 0.009 \\
55376.420 & 16.110 & 0.012 \\
55380.467 & 15.931 & 0.009 \\
55381.405 & 16.090 & 0.009 \\
55381.455 & 16.062 & 0.008 \\
55386.397 & 15.940 & 0.007 \\
55386.444 & 15.933 & 0.006 \\
55391.400 & 15.662 & 0.013 \\
55391.444 & 15.705 & 0.011 \\
55396.411 & 15.123 & 0.007 \\
55396.466 & 14.971 & 0.008 \\
55401.304 & 14.843 & 0.012 \\
55401.348 & 14.822 & 0.013 \\
55407.294 & 14.261 & 0.013 \\
55407.342 & 14.263 & 0.011 \\
55410.295 & 14.057 & 0.007 \\
55410.340 & 14.015 & 0.009 \\
55413.292 & 13.893 & 0.008 \\
55413.336 & 13.886 & 0.006 \\
55416.287 & 13.755 & 0.009 \\
55416.330 & 13.774 & 0.013 \\
55419.282 & 13.568 & 0.009 \\
55419.326 & 13.567 & 0.011 \\
55426.161 & $<$13.409 & ... \\
55429.177 & $<$13.444 & ... \\
55429.220 & $<$13.420 & ... \\
55435.303 & $<$13.297 & ... \\
55438.330 & $<$13.124 & ... \\
55438.413 & $<$13.119 & ... \\
55441.365 & $<$13.231 & ... \\
55441.430 & $<$13.201 & ... \\
55444.390 & $<$13.084 & ... \\
55444.434 & $<$13.041 & ... \\
55448.144 & $<$13.017 & ... \\
55448.189 & $<$12.990 & ... \\
55451.156 & $<$13.134 & ... \\
55454.265 & $<$13.104 & ... \\
55460.339 & $<$13.190 & ... \\
55460.388 & $<$13.159 & ... \\
55463.341 & $<$13.137 & ... \\
55463.386 & $<$13.128 & ... \\
55466.326 & $<$13.112 & ... \\
55477.191 & $<$13.173 & ... \\
55477.234 & $<$13.157 & ... \\
55484.263 & $<$13.166 & ... \\
55485.119 & $<$13.260 & ... \\
55485.163 & $<$13.299 & ... \\
55497.098 & $<$13.233 & ... \\
55497.141 & $<$13.232 & ... \\

\hline
\end{tabular}
}
\label{tab-P48}
%\end{center}
\end{minipage}
\end{table}

% \enddata
% \end{deluxetable}

\end{document}